\documentclass[preprint,aps]{revtex4-1}
\usepackage{graphicx}
\usepackage{rotating}
\usepackage{bm}
\usepackage{subfigure}

\begin{document}
\title{Many body density of states in the edge of the spectrum: non-interacting limit}
\author{Pragya Shukla}
\affiliation{Department of Physics, Indian Institute of Technology, Kharagpur-721302, India  }
\date{\today}

\widetext

\begin{abstract}
 
In noninteracting limit, the density of states of a many body system can be expressed as the convolution of single body density of states of its subunits. Here we use the formulation to derive the ensemble averaged many body density of states for the cases in which subunits can be modelled by Gaussian or Wishart random matrix ensembles.

\end{abstract}

\maketitle

\section{Introduction}

I would like to dedicate this work to the meomory of Fritz Haake, to celebrate his rich and diverse contributions to Mathematical Physics, especially Quantum Chaos. It was his book \cite{fh} on the latter topic, an important resource for  my graduate studies and I am deeply indebted to him for writing it.     

Consider a many body system composed of independent or weakly interacting  particles with mutual interaction not strong enough to perturb the single particle spectrum. The latter situation may arise, for example, 
for the systems in which the interactions are local within a short spatial scale, say $L_s$ and negligible at large scales. The system can then be described as a collection of non-interacting/ weakly interacting subunits of linear size $L_s$ and its Hamiltonian can  be expressed as a tensor sum of the local Hamiltonians (of the subunits) and its physical properties can then be determined from those of subunits (referred as ''particles'' hereafter).  The many body density of states ({\it mdos}) can then be obtained by a convolution of the single particle states ({\it sdos}) \cite{brody}.

For many body systems with non-local interactions, the bulk density of many body states is very large and even a small non-local interaction can in general mix the single particle states, wiping out their details (due to small level spacing). The convolution route to derive {\it mdos} is therefore  not valid in the bulk of the spectrum of such a system.  The situation however is different in their edge: due to vanishing {\it dos} in the edge,  the separation between levels is large and the many body interactions may not be strong enough to mix the single particle energy states. The {\it mdos} in the edge of the spectrum can then be derived by the convolution of single particle spectrum.

For a system consisting of a large number of particles and with a very weak restriction on {\it sdos}, the convolution route predicts, based on central limit theorem (CLT), a Gaussian behaviour of the {\it mdos} in the bulk of the spectrum \cite{brody}. The CLT however is not applicable in the regime where the single particle spectrum is singular e.g. the spectral-edge. With growing interest in the low temperature physics, it becomes now important to know the behaviour of the {\it mdos} near the edge i.e the region where the ground state of the many body system lies. This motivates us to  consider following questions: 

(i) assuming the theoretical form  for the {\it sdos} in the single body spectral edge is known, what would be the behaviour of {\it mdos} in the same energy range, 

(ii) for what cases of {\it sdos}, the {\it mdos} will retain the same functional form albeit approximately?

The standard route to determine the energy states  of a system is based on a matrix representation of the Hamiltonian in a physically motivated basis. Based on the nature of interactions between different subunits, referred here as "particles", the Hamiltonian matrix can be of various types e.g. dense  (full), sparse, banded and subjected to additional matrix constrains due to symmetry and conservation laws \cite{fh,me,psijmp}. Besides, the complexity of interactions often renders an exact  calculation of  the matrix elements a technically challenging task. The deterministic error appears as a distribution of the matrix element around its most probable value and  the Hamiltonian matrix behaves as a random matrix with some or all random elements.  The physical properties of the Hamiltonian matrix are then best described by an appropriate  ensemble of its replicas, with ensemble parameters determined from the system conditions. With density of states as a backbone of many physical properties, it is therefore  necessary to derive its general formulation applicable for a wide range of systems. This is however technically very complicated due to  system dependence of the randomization of the many body Hamiltonian.  The {\it mdos} also varies from one replica to the other, thus making it necessary to consider its average as well fluctuations across the ensemble.
However for cases in which {\it sdos} is known apriori, the {\it mdos} can be 
determined through convolution route.

It is now well-established, theoretically as well as experimentally, that the fluctuation properties of a  one body (single particle) operator in the ergodic regime of its wavefunction dynamics can be well described by the Wigner-Dyson ensembles  of Hermitian matrices if the operator is Hamiltonian and by stationary Wishart ensembles if it is a semi-positive definite (can be described by a  full Wishart random matrix) \cite{fh,me,psijmp}.  
In case of the localized/ partially localized  wavefunctions or weakly violated exact symmetries of one body operator, however, the statistics is described by generalized ensembles e.g. multi-parameter dependent sparse random matrix ensembles, brownian ensembles etc \cite{psijmp, pswf, psand, psall}. The present study analyses the ensemble averaged  {\it mdos} in the spectrum edge of a many particle system that consists of many non-interacting particles, with their fluctuation properties described by Gaussian or Wishart ensembles of both stationary/ non-stationary types \cite{sp,pslg}.

The paper is organized as follows. Our approach is based on first writing the many body density of states {\it mdos} as the convolution of single body densities {\it sdos}. For the systems consisting of smaller subunits  e.g. a $2$-body system with known {\it sdos}, the {\it mdos}  can be obtained directly from the convolution integral. However the integration route becomes increasingly complicated as the number of subunits increase, motivating us ti consider an alternative route based on the  differential equation satisfied by the {\it sdos}.  A repeated application of the latter to convolution integral leads to a non-homogeneous differential equation for {\it mdos}  which can then be solved to obtain the {\it mdos}, subjected to the boundary conditions derived from a smooth connection of the edge {\it mdos} with that of bulk.  Under some approximations,  {\it mdos} can also be given by the homogeneous solution of the related differential equation.

\section{Average many body density of states} 

Consider the case in which the system Hamiltonian $H$ with $g$ interacting subunits can be written as
\begin{eqnarray}
H=H_0 +V
\label{hi} 
\end{eqnarray}
with $H_0$ as a sum over non-interacting Hamiltonians ${\mathcal H}^{(s)}$, $s=1 \to g$, 
\begin{eqnarray}
H_0 = \sum_{s=1}^g \;   {\mathcal H}^{(s)}
\label{h0} 
\end{eqnarray}
and $V$ as the many body interaction among subunits. Here we consider the energy limit in which $V$ can be ignored e.g. a system generated by distributing $g$ non-interacting particles over $N$ single particle states ${\mathcal E}_k$ with $k=1 \to N$ (ignoring the occupancy constraints due to Pauli principle for cases $g \ll N$, referred as dilute limit). Another energy range, where $V$ often plays a negligible role, is the lower edge region of the many body spectrum and is the focus of current study.

Assuming $E_1, E_2, \ldots, E_M$ as the eigenvalues of $H$, with $M=N^g$ as the size of $H$-matrix, the many body {\it dos} $\rho_g(e)$  can be expressed as 
\begin{eqnarray}
\rho_g(e) =  \sum_{n=1}^M \delta(e-E_n)
\label{rho}
\end{eqnarray}
Using the same formulation for the single particle {\it dos} $\rho_1({\mathcal E})$,  (with ${\mathcal E}_1, {\mathcal E}_2, \ldots, {\mathcal E}_M$ as the eigenvalues of ${\mathcal H}$), we have 
\begin{eqnarray}
\rho_1({\mathcal E}) =\sum_{k=1}^N \delta({\mathcal E}-{\mathcal E}_k)
\label{rho}
\end{eqnarray}

 In non-interacting limit, the energy $E_n$ of a many body levels $\Psi_n$ can further be expressed in terms of the single single particle state as follows.  Using  $\hat{n_k}$ as the number operator for the $k^{th}$ single particle state ${\mathcal E}_k$ (counting the number of particles with single particle energy ${\mathcal E}_k$ in the manybody state $\Psi_n$), $H$ represented in the $N^g$-dimensional product basis of single particle states becomes
\begin{eqnarray}
H = \sum_k  \hat{n_k}  \; {\mathcal E}_k
\label{hh}
\end{eqnarray}
This in turn leads to $E_n = \langle \Psi_n | H \Psi_n \rangle = \sum_k n_k \;  {\mathcal E}_k$ where $n_k = \langle \Psi_n | \hat{n_k} |\Psi_n \rangle$

Following from the above, $\rho_g(e)$ can then be expressed as the  covolution of many body {\it dos} of $(g-1)$ particles and a single particle {\it dos}
\begin{eqnarray}
\rho_{g}(e) &=& \int_{e_a}^{e_b} \rho_{g-1}(x) \; \rho_{1}(e-x) \; {\rm d}x\label{mbd}
\end{eqnarray}
with notation $\rho_m$ denoting the {\it dos} of $m$ particles and $e_a, e_b$ as the many body energy range of interest.

An averaging of $\rho_g(e)$ over all replicas of $H$ in the ensemble at a fixed $e$ then leads to its ensemble average at $e$: 
\begin{eqnarray} 
 \langle \rho_g(e) \rangle = \sum_{n=1}^M \langle \delta(e-E_n) \rangle. 
\label{rhoe}
\end{eqnarray}
As  the particles are  almost mutually  independent, an
ensemble averaging of eq.(\ref{mbd}) can be expressed as
\begin{eqnarray}
\langle \rho_{g}(e) \rangle =\int_{e_a}^{e_b} \langle  \rho_{g-1}(x) \rangle \; \langle \rho_{1}(e-x) \rangle\; {\rm d}x
\label{mbd1}
\end{eqnarray}
For later purposes, we also consider an alternative formulation  for even $g$ e.g. $g=2^n$, $n=1,2, \ldots$;  $\rho_{2^n}(e)$ can be rewritten as 
\begin{eqnarray}
\langle \rho_{2^n}(e) \rangle =\int_{e_a}^{e_b} \; \langle \rho_{2^{n-1}}(x) \rangle\; \langle \rho_{2^{n-1}}(e-x) \rangle\; {\rm d}x.
\label{rhoea}
\end{eqnarray}
The above in turn leads to 
\begin{eqnarray}
\langle \rho_{2^n+1}(e) \rangle =\int_{e_a}^{e_b} \; \langle \rho_{2^{n}}(x) \rangle\; \langle \rho_{1}(e-x) \rangle\; {\rm d}x.
\label{rhooa}
\end{eqnarray}

Hereafter our analysis is confined to the edge region of the single body spectrum. Using the notation $\langle \rho_{m, edge}(e) \rangle$  for the ensemble averaged $m$-body {\it dos} in the edge region of the spectrum, 
 we have, from eq.(\ref{mbd1}),
\begin{eqnarray}
\langle \rho_{g, edge}(e) \rangle =\int_{-\infty}^{e_0} \langle  \rho_{g-1, edge}(x) \rangle \; \langle \rho_{1,edge}(e-x) \rangle\; {\rm d}x
\label{mbd1a}
\end{eqnarray}
with $-\infty < e < e_0$ as the single body spectral edge. Here we note that due to rapidly increasing {\it mdos}, the many body spectral edge indeed lies below single body spectral edge $e_0$. Although the sum over contributions of single particle spectral edges of $g$ bodies is $g \; e_0$, the number of many body levels near this energy is  quite large (e.g $\sim g$), thus making it part of the bulk spectrum instead of the edge.

Further mathematical steps depend on the behaviour of  $\langle \rho_{1,edge}(e-x) \rangle$ which can vary based on the complexity of the system. As mentioned in previous section, here we confine the analysis to the many body systems consisting of   single body Hamiltonians which  can be modelled by Gaussian and Wishart ensembles. Moreover, with present study  focussed on the edge behaviour of the average spectral density $\rho_{g, edge}$, hereafter the subscript $''edge''$ will be suppressed unless needed for clarity of presentation.

\section{Stationary Gaussian ensembles}

A stationary Gaussian ensemble, also known as Wigner-Dyson ensemble, consists of Hermitian matrices (denoted here by symbol ${\mathcal H}$) with basis-invariant Gaussian ensemble density  $\rho_{\mathcal H} ({\mathcal H}) \propto {\rm exp}[{-{\rm Tr} {\mathcal H}^2 }]$. The underlying symmetry constraints lead to three universality classes of WDEs, labelled by the symmetry parameter $\beta$: (i) Gaussian orthogonal ensembles (GOE) of real-symmetric matrices ($\beta=1$), (ii) Gaussian unitary ensembles (GUE) of complex Hermitian matrices ($\beta=2$), (iii) Gaussian symplectic ensembles (GSE) of real-quaternion matrices ($\beta=4$). The average {\it dos} for a Wigner-Dyson ensemble follows a  semicircle behaviour in the bulk of the spectrum and an Airy function type behaviour in its edge.  

Following standard notation, the ensemble averaged {\it sdos} is given as  $\langle \rho_{1}(e) \rangle =K_N(e,e)$  where the kernel $K$ has different scaling behaviour in the bulk  and edge of the spectrum.
For a stationary Gaussian ensemble, it has been shown to behave as follows \cite{fo1}: $\lim_{N \to \infty} {1\over \sqrt{2N}} \; K_N(e\sqrt{2N}, e\sqrt{2N})={1\over \pi} \sqrt{1-e^2}$  in bulk ($e \sim 0$)  and $$\lim_{N \to \infty} {1\over 2^{1/2} N^{1/6}} K_N \left(-\sqrt{2N}+{e\over 2^{1/2} N^{1/6}}, -\sqrt{2N} +{e\over 2^{1/2} N^{1/6} }\right) =f_1(e)$$ in the lower edge ($e \sim -\sqrt{2N}$) where $f_1(e)$ depends on the symmetry class of the ensemble.

Shifting the origin of spectrum to $e=-\sqrt{2N}$, the {\it sdos} $\langle \rho_{1}(e) \rangle$  in the lower edge region, for a single particle Hamiltonian modelled by a stationary Gaussian ensemble, can then be given as
\begin{eqnarray}
\langle \rho_{1}(e) \rangle = { \gamma \; N} \; \; 
f_1( \gamma \;e) \hspace{0.1in} - \infty  < e < e_0. 
\label{mbd0}
\end{eqnarray}
with $\gamma=2^{1/2} N^{1/6}$.

As our approach is based on the convolution of {\it sdos}, it is useful to define the {\it mdos} also in a rescaled form 
\begin{eqnarray}
\langle \rho_{g}(e) \rangle= \gamma \; N^g \; f_g( \gamma e).
\label{rhogg}
\end{eqnarray}
From eq.(\ref{mbd1a}), we then have 
\begin{eqnarray}
f_g(e) =\int_{-\infty}^{e_0} f_{g-1}(x)  \; f_{1}(e-x) \rangle\; {\rm d}x
\label{fg}
\end{eqnarray}
Similarly eq.(\ref{rhoea}) gives 
\begin{eqnarray}
f_{2^n}(e) =\int_{-\infty}^{e_0} f_{2^{(n-1)}}(x)  \; f_{2^{(n-1)}}(e-x) \rangle\; {\rm d}x
\label{feven}
\end{eqnarray}
The fom of $f_1(e)$ depends on the symmetry class of the ensemble and can thereby lead to different $f_g(e)$. Here we  consider the cases with $f_1$ given by a GOE or GUE only. The steps can however be extended directly to GSE too. 

\subsection{Gaussian Unitary ensemble (GUE)}

For a GUE, $f_1(e)$ in the lower edge is
\begin{eqnarray}
f_1(e) & \approx & e \; {\rm Ai}^2(-e)  + ({\rm Ai}'(-e))^2  
\label{fnt}  
\end{eqnarray}
with ${\rm Ai}(e)$ and ${\rm Ai}'(e)$ as the Airy function of the first kind and its derivative, respectively. We note that the exact differential equation satisfied by  $f(y)$  is \cite{fo1}
\begin{eqnarray}
 \left({{\rm d}^3  \over {\rm d}y^3} + 4 \; y \; {{\rm d}  \over {\rm d}y} - 2 \; \right) f_1(y) =0
\label{fnt1}
\end{eqnarray}

To derive $\langle \rho_{g}(e) \rangle$ in the regime $e < e_0$, we proceed  iteratively and start with $g=2$. 
Using the definition 
\begin{eqnarray}
\langle \rho_{g}(e) \rangle= \gamma \; N^g \; f_g( \gamma e)
\label{rhogg}
\end{eqnarray}
 with $f_1(x)$ given by eq.(\ref{fnt}), a substitution of eq.(\ref{mbd0}) in eq.(\ref{mbd1}) gives $f_2$ (the rescaled $2$-body {\it dos})  in the edge  as 
\begin{eqnarray}
f_{2}(y) =\int_{-\infty}^{e_0} \; f_{1}(x)\; f_{1}(y-x) \; {\rm d}x.
\label{rho2a}
\end{eqnarray}
It is easy to check that $f_{2}(y)$ satisfies 
\begin{eqnarray}
\left({{\rm d}^3  \over {\rm d}y^3} + 4 \; y \; {{\rm d}  \over {\rm d}y} - 2 \; \right)\; f_2(y) =- 4 \; I_2(y).
\label{mbd2a}
\end{eqnarray}
with
\begin{eqnarray}
I_2(y) = \int_{-\infty}^{e_0} \;x \; f_1(x) \; {\partial f_{1}(y-x)  \over \partial x} \; {\rm d}x. 
\label{i2a}
\end{eqnarray}
where we have used the relation ${\partial f_1(y-x) \over \partial (y-x)} =- {\partial f_1(y-x) \over \partial x} =  {\partial f_1(y-x) \over \partial y}$.
Applying partial integration, $I_2$ can now be rewritten as
\begin{eqnarray}
I_2(y) = 
-{3\over 2} \; f_2(y)  +{1 \over 4} \; {{\rm d}^2 f_2(y)  \over {\rm d}y^2} +Q_2(y,e_0)
\label{i3}
\end{eqnarray} 
with 
\begin{eqnarray}
Q_2(y,x) &=& {1\over 4} \left[ f_1(y-x) \; {\rm D}_x^2 f_1(x) -({\rm D}_x f_1(y-x))  ({\rm D}_x f_1(x)) + ({\rm D}_x^2 f_1(y-x)) f_1(x)\right] + \nonumber \\
& + & x \; f_1(x) \; f_1(y-x)
\label{q2}
\end{eqnarray} 
where $D^k_x \equiv {{\rm d}^k\over {\rm d}x^k}$.

Substitution of the above in eq.(\ref{mbd2a}) gives 
\begin{eqnarray}
\left( 2 {{\rm d}^3  \over {\rm d}y^3} + 4 \; y \; {{\rm d}  \over {\rm d}y} - 8 \; \right) f_2(y) = - 4 \; Q_2(y, e_0) 
\label{mbd3}
\end{eqnarray}

Further using eq.(\ref{feven}) with $n=2$ and applying the differential operator $\left(2 {{\rm d}^3  \over {\rm d}y^3} +4 \; y \; {{\rm d}  \over {\rm d}y} -8  \; \right)$ to both sides of the integral then leads to a differential equation for $f_{4}(e)$. Repeating the steps multiple times with $n=3,...m$ then  leads to the differential equation   for $f_{g}(e)$ with $g=2^n$, 

 \begin{eqnarray}
\left( a_{g} \; {{\rm d}^3  \over {\rm d}y^3} + 4 \; y \; {{\rm d}  \over {\rm d}y} - b_{g} \; \right) \; f_{g}(y) = - 4 \;  Q_{g}(y,e_0) -4 \; G_{g}(y,e_0)
\label{mbd3g}
\end{eqnarray}
where $a_{2^n} =2^n$, $b_{2^n}= 2 (2+b_{2^{(n-1)}})$ with $a_1=1, b_1=-1$ for $n=0$) and 
\begin{eqnarray}
 G_{2^n}(y,e_0) = \int  \left[Q_{2^{n-1}}(y-x) + G_{2^{n-1}}(y-x) \right] \; f_{2^{n-1}}(x)r
 \; {\rm d} x 
\label{gn}
\end{eqnarray}
and
\begin{eqnarray}
Q_{2^n}(y,x) &=& {2^{(n-1)}\over 4} \left[ f_{2^{(n-1)}}(y-x) \; {\rm D}_x^2 f_{2^{(n-1)}}(x) -({\rm D}_x f_{2^{(n-1)}}(y-x))  ({\rm D}_x f_{2^{(n-1)}}(x))  \right . + \nonumber \\
&+&  \left . ({\rm D}_x^2 f_{2^{(n-1)}}(y-x)) f_{2^{(n-1)}}(x) \right]+ x \; f_{2^{(n-1)}}(x) \; f_{2^{(n-1)}}(y-x)
\label{q1}
\end{eqnarray}

Using eq.(\ref{mbd3g}) with $g=2^n$ and substitution in the definition 
$f_{2^n+1}(y) =\int_{-\infty}^{e_0} \; f_{2^n}(x)\; f_{1}(y-x) \; {\rm d}x$ followed by partial integration and rearrangements, we can derive the  differential equation for $f_{2^n+1}(e) $; the latter turns out to be same as eq.(\ref{mbd3g}) but now $g=2^n +1$, $a_{2^n+1} =2^n+1$, $b_{2^n+1}= 2 (5+b_{2^n})$ and
\begin{eqnarray}
 G_{2^n+1}(y,e_0) = \int  \left[Q_{2^n}(y-x) + G_{2^n}(y-x) \right] \; f_1(x) \; {\rm d} x 
\label{gno}
\end{eqnarray}
and
\begin{eqnarray}
Q_{2^n+1}(y,x) &=& {2^n \over 4} \left[ f_{2^n}(x) \; {\rm D}_x^2 f_1(y-x) -({\rm D}_x f_{2^n}(y-x))  ({\rm D}_x f_{1}(x)) \right . + \nonumber \\&+& \left . ({\rm D}_x^2 f_{2^n}(y-x)) f_{1}(x) \right]+ x \; f_{2^n}(x) \; f_{1}(y-x)
\label{q2}
\end{eqnarray}

Eq.(\ref{mbd3g}) is a linear nonhomogeneous differential equation of the $3^{rd}$ order. Writing its  general solution as a sum of homogeneous and nonhomogeneous parts 
\begin{eqnarray}
f_g(y) = f_{g,h}(y) +  f_{g,ih}(y),
\label{fgh}
\end{eqnarray}
the three independent solutions of the homogeneous part $f_{g,h}(y)$ can be given as 
\begin{eqnarray}
 f_{g,h}(y)  &=& c_1 \; F\left(-{b_g\over 12},  \{{1\over 3},{2\over 3}\}, {- 4 y^3\over 9 a_g} \right) + c_2 \; \left( {4 y^3\over 9 a_g}\right)^{1/3} \; F\left(\{{1\over 3}-{b_g\over 12}\},  \{{2\over 3},{4\over 3}\}, {- 4 y^3\over 9 a_g}\right) \nonumber \\
 &+& c_3 \; \left( {4 y^3\over 9 a_g}\right)^{2/3} \; F\left({2\over 3}-{b_g \over 12},  \{{4\over 3},{5\over 3}\}, {- 4 y^3\over 9 a_g}\right)
\label{mbd3h}
\end{eqnarray}

As a check, we note that, for $g=1$, eq.(\ref{mbd3g}) reduces to linear homogeneous differential equation given by eq.(\ref{fnt1}) and the solution given by eq.(\ref{mbd3h}) reduces to eq.(\ref{fnt}) for $c_1=0.066987, c_2=0.091888095, c_3=0.12604490$. Clearly $f_{1,ih}=0$.  For arbitrary $g > 1$ case, however, it is apriori not obvious whether $f_{g,ih}$ can be ignored. 

For large $g$ cases, which are of wider physical interests, here we consider an altrenative although approximate route: it is based on neglecting the first order derivative in eq.(\ref{mbd3g}), its coefficient being much smaller than the coefficients of the other two terms on the left of eq.(\ref{mbd3g}). The approximation reduces the latter equation to a linear differential equation with 
constant coefficients:    
\begin{eqnarray}
\left( a_{g} \; {{\rm d}^3  \over {\rm d}y^3} - b_{g} \; \right) \; f_{g}(y) = - 4 \;  Q_{g}(y,e_0) -4 \; G_{g}(y,e_0)
\label{mbd3k}
\end{eqnarray}
and its solution can then be given as 
 \begin{eqnarray}
 f_{g}(y)  = \sum_{k=1}^3 c_k \; {\rm e}^{t_k y} \left[1+ A_k \; \int    {\rm e}^{t_k x} \left(Q_g(x) + G_g(x) \right) \right]
\label{mbd3l}
\end{eqnarray}
with $t_k$ as the roots of equation $(D^3 - b_g/a_g) =0$ and $A_k= {1\over 3 \;  t_k^2}$. Here two of the arbitrary constants $c_1, c_2, c_3$ can be determined by  imposing the requirement that $f_g(-\infty) \to 0$ and  $f_{g}(e_0)$ approaches correct bulk density smoothly.
Substitution of eq.(\ref{mbd3h}) in eq.(\ref{rhogg}) then leads to {\it mdos}.
The remaining constant can be determined by invoking the normalization condition on full {\it mdos}. Here we note that,
for a {\it sdos} modeled by a GUE, the {\t mdos} in the bulk turns out to be a Gaussian or semicircle based on whether the nature of partice interactions \cite{brody}.

\subsection{Gaussian orthogonal ensemble (GOE)}

As in the previous case,  the {\it sdos} $\langle \rho_{1}(e) \rangle$  in the lower edge region, for a single particle Hamiltonian modeled by a GOE, can again be given by eq.(\ref{mbd0}) but now  
\begin{eqnarray}
f_1(x) & \approx & x \; {\rm Ai}^2(-x)  + ({\rm Ai}'(-x))^2 + {1 \over 2} \;  {\rm Ai}(-x) \; \int_{-\infty}^{-x} {\rm Ai}(y) \; {\rm d}y.  
\label{ft}  
\end{eqnarray}
with ${\rm Ai}(x)$ same as in eq.(\ref{fnt}).
We note that the origin of spectrum here again is shifted to  $e=-\sqrt{2N}$.
 It is easy to check that $f_{1}(y)$ satisfies 
\begin{eqnarray}
 \left( {{\rm d}^2  \over {\rm d}y^2}  +y \; \right) \; f_{1}(y)  = Q_1(y)
\label{dwg}
\end{eqnarray}
with
\begin{eqnarray}
 Q_1(y) = y^2 \; {\rm Ai}^2(-y)  + y ({\rm Ai}'(-y))^2 -
 {1\over 2} \; {\rm Ai}(-y) {\rm Ai}'(-y)
\label{dwg}
\end{eqnarray}

To derive $\langle \rho_{g}(e) \rangle$ in the edge regime $e < e_{0}$, we again use relation (\ref{rhogg}), proceed  iteratively and use eq.(\ref{mbd1}) with $g=2$, equivalently eq.(\ref{rho2a}). A double differentiation of the latter along with rearrangement of the terms then leads to following differential equation for $f_2$

\begin{eqnarray}
\left({{\rm d}^2  \over {\rm d}y^2} + y \right)\; f_{2}(y) = I_2(y) + Q_2(y).
\label{mbd2}
\end{eqnarray}
where 
\begin{eqnarray}
 I_2(y) &=&\int_{-\infty}^{e_0} {{\rm d}^2 f_{1}(x)  \over {\rm d}x^2}  \;  f_{1}(y-x)  \; {\rm d}x\\
Q_2(y) &=&\int_{-\infty}^{e_0} \; \left(\; Q_1(x) \; f_{1}(y-x) + f_{1}(x)  \; Q_1(y-x)\right) \; {\rm d}x. 
\label{i2}
\end{eqnarray}
Applying repeated partial integration, we have 
\begin{eqnarray}
I_2(y) =u_2(y) -{{\rm d}^2 f_{2}(x)  \over {\rm d}y^2}
\label{i2}
\end{eqnarray}
where $u_2(y) =  f_{1}(e_0) {{\rm d} f_{1}(y-e_0)  \over {\rm d}e_0} - {{\rm d} f_{1}(e_0)  \over {\rm d}e_0}  \; f_{1}(y-e_0)$.
Substitution of the above in eq.(\ref{mbd2}) reduces it as
\begin{eqnarray}
 \left(2 \; {{\rm d}^2  \over {\rm d}y^2} + y \right)\; f_{2} = u_2(y) + Q_2(y)
\label{mbd2x}
\end{eqnarray}
Further using eq.(\ref{rhoea})  and  repeating the above steps then leads to the differential equation for $f_{2^n}(y)$. The latter along with eq.(\ref{rhooa})  can subsequently be used to derive the differential equation for $f_{2^n+1}$. For both cases ($g=2^n$ or $g=2^n+1$), we have  
 \begin{eqnarray}
\left[g \; {{\rm d}^2  \over {\rm d}y^2} + y \right] \; f_{g}(y)  = u_{g}(y) + Q_{g}(y).
\label{mbd4}
\end{eqnarray}
where 
\begin{eqnarray}
u_{g}(y) = \sum_{m=1}^{g/2} \left[ {{\rm d} f_{g-m}(e_0)  \over {\rm d}e_0}  \;  f_{m}(y-e_0)  -f_{g-m}(e_0) {{\rm d} f_{m}(y-e_0) \over {\rm d}e_0}\right]
\label{un}
\end{eqnarray}
and
\begin{eqnarray}
Q_{2^n}(y) &=&\int_{-\infty}^{e_0} \; \left[\; Q_{2^{n-1}}(x) \;f_{2^{n-1}}(y-x) + f_{2^{n-1}}(x)  \; Q_{2^{n-1}}(y-x)\right] \; {\rm d}x,\\
Q_{2^n+1}(y) &=& \int_{-\infty}^{e_0} \; \left[\; Q_{2^{n}}(x) \; f_{1}(y-x) + f_{2^n}(x)  \; Q_{1}(y-x)\right] \; {\rm d}x
 \label{qn}
\end{eqnarray}

Eq.(\ref{mbd4}) is exact and its solution can be written as  $f_g(y)= f_{g,h}(y) +  f_{g,ih}(y)$ with $f_{g,h}$ and $f_{g,ih}$ as its homogeneous and nonhomogeneous solutions. Here the two independent homogeneous solutions are
\begin{eqnarray}
f_{g,h}(y) = c_1 \; {\rm Ai}(- g^{-1/3}\; y) +c_2 \; {\rm Bi}(- g^{-1/3} \; y)
\label{mbd4h}
\end{eqnarray}
with ${\rm Ai}(y)$ same as in eq.(\ref{fnt}) and $Bi(y)$ as the Airy functions of the second kind. Further, as in the GUE case,  the constants $c_1, c_2$ can be determined from the boundary conditions on $f_{g}(y)$, namely, $f_g(-\infty) \to 0$ and  $f_{g}(e_0)$ approaches correct bulk density smoothly.
Using the standard approach (based on the wronskin of two independent homogeneous solutions of a second order differential equation), the inhomogeneous solution of eq.(\ref{mbd4}) is then 
\begin{eqnarray}
f_{g,ih}(y) &=& A(y) \; {\rm Ai}(- g^{-1/3}\; y) +B(y) \; {\rm Bi}(- g^{-1/3} \; y).
\label{mbd4i}
\end{eqnarray}
with
\begin{eqnarray}
A(y) &=& \int^y  (w(x))^{-1} \; {{\rm Bi}(- g^{-1/3}\; x) \; \left[u_{g}(x) + Q_{g}(x)\right]} \;  {\rm d}x, \\  
B(y) &=& \int^y (w(x))^{-1} \;{ {\rm Ai}(-g^{-1/3}\; x) \; \left[u_{g}(x) + Q_{g}(x) \right]} \; {\rm d}x.
\label{ab}
\end{eqnarray}
The above integrals can be caculated by a substitution of eq.(\ref{un}) and eq.(\ref{qn}). Here the wronskin $w(x)={\rm Ai}(- g^{-1/3}\; x) {\rm Bi}'(- g^{-1/3}\; x) - {\rm Bi}(- g^{-1/3}\; x){\rm Ai}'(- g^{-1/3}\; x)$ turns out to be a constant.
Further addition of eq.(\ref{mbd4h}) and eq.(\ref{mbd4i}) and subsequently imposing the boundary condition $f_g(-\infty)=0$ then leads to   $f_g(y) \approx (c_1+A(y)) \; {\rm Ai}(-g^{-1/3}\; y) $. The latter along with eq.(\ref{rhogg}) then leads to ensemble averaged {\it mdos} 
\begin{eqnarray}
\langle \rho_g(e)\rangle \approx (c_1+A(y)) \; \gamma \; N^g \; {\rm Ai}(-g^{-1/3}\; \gamma \; e) 
\label{fo3}
\end{eqnarray}

\section { Stationary Wishart ensembles (WE)}

A stationary Wishart ensemble consists of Hermitian matrices  $L \equiv C^{\dagger}. C$ with $C$ as a $(N+\alpha) \times N$ complex matrix. The ensemble averaged {\it sdos} for this case can again be given as  $\langle \rho_{1}(e) \rangle =K_N(e,e)$  where the kernel $K$ has different scaling behavior in the bulk  and edge of the spectrum.
The kernal $K$ for average single particle {\it dos} in this case has following scaling behavior: $\lim_{N \to \infty} {1\over \sqrt{2N}} \; K_N(e/2N, e/2N)={1\over 2\pi  \sqrt{e}} \sqrt{1-e}$  in bulk   and $\lim_{N \to \infty} {1\over 4N} K_N \left({e\over 4N}, {e\over 4N }\right) =f_1(e)$ in the lower edge ($e \sim 0$) where $f_1(e)$ again depends on the symmetry class of the Wishart ensemble \cite{fo1,fo2, tf}.

Following from the above, the {\it sdos} $\langle \rho_{1}(e) \rangle$  in the lower edge region, for a single particle operator modeled by a stationary Wishart esemble, can then be given as
\begin{eqnarray}
\langle \rho_{1}(e) \rangle = { \gamma \; N} \; \; 
f_1( \gamma \;e) \hspace{0.1in} 0  < e < e_0. 
\label{rho1w}
\end{eqnarray}
with $\gamma=4 N$.
Here again the scaled form of {\it mdos} is given by eq.(\ref{rhogg}) but with $\gamma$ as in eq.(\ref{rho1w}).

\subsection { Wishart Unitray ensemble (WUE)}

For a single particle operator modeled by a WUE, $f_1(x)$ in the edge region can  be given as (eq.(2.10) of \cite{fo1})
\begin{eqnarray}
f_1(x) =
{1\over 4} \left[ (J'_{a}(\sqrt{x})^2 + J_a(\sqrt{x}) J'_{a+1}(\sqrt{x}) -{a \over \sqrt{x}} J_a(\sqrt{x}) J_a'(\sqrt{x}) +{1 \over \sqrt{x}} J_a(\sqrt{x}) J_{a+1}(\sqrt{x})\right]  \nonumber \\
\label{fw}
\end{eqnarray}

As discussed in  \cite{fo1},   $f_{1}(y)$ in this case satisfies 
\begin{eqnarray}
 \left( y \;{{\rm d}^2  \over {\rm d}y^2} -\alpha \; {{\rm d}  \over {\rm d}y} - 1\right)  f_{1}(y) =  {\alpha \over y} \; f_{1}(y)
\label{dw}
\end{eqnarray}

Again using eq.(\ref{mbd1}) and proceeding as in previous cases, it can be shown that $\langle \rho_{2}(y) \rangle$ now satisfies 
\begin{eqnarray}
\left( y \;{{\rm d}^2  \over {\rm d}y^2} -\alpha \; {{\rm d}  \over {\rm d}y} - 1\right) \; f_{2}(y)  = I_2(y) - \alpha \; h_2(y).
\label{mbdw}
\end{eqnarray}
where now
\begin{eqnarray}
I_2(y) =\int_0^{e_0} \;x \; f_1(x) \; {\partial^2 f_1(y-x) \over \partial x^2}\; {\rm d}x. 
\label{iw2}
\end{eqnarray}
and
\begin{eqnarray}
h_2(y) =\int_0^{e_0} { f_1(x) \;f_1(y-x)  \over y-x}\; {\rm d}x. 
\label{hy}
\end{eqnarray}

Applying partial  integration, $I_2$ can be rewritten as 
\begin{eqnarray}
I_2(y) = {Q_2\over 2} + {y \over 2} \; {{\rm d}^2  \over {\rm d}y^2} \; f_2(y) 
\label{iw3}
\end{eqnarray}
Substitution of the above in eq.(\ref{mbdw}) leads to
\begin{eqnarray}
\left( {y\over2} \;{{\rm d}^2  \over {\rm d}y^2} -\alpha \; {{\rm d}  \over {\rm d}y} - 1 \right) f_2(y)  = {Q_2\over 2}- \alpha \; h_2(y)
\label{iw4}
\end{eqnarray}
where 
\begin{eqnarray}
 Q_2(y)=b \left(f_1(b)  {\partial f_1(y-b)   \over \partial b} -  f_1(y-b)  {\partial f_1(b)  \over \partial b}  \right) - f_1(b) f_1(y-b) 
\label{iw5}
\end{eqnarray}

Again using eq.(\ref{rhogg}) for $g=4, 8,\ldots, 2^n$, $n$ arbitrary, and iterating the above steps repeatedly (as in previous cases) leads to 
\begin{eqnarray}
 \left({y \over 2^n} \; {{\rm d}^2  \over {\rm d}y^2} - \alpha \; {{\rm d}  \over {\rm d}y} - 1 \; \right) \; f_{2^n}(y)  = R_{2^n}(y)
\label{dwg}
\end{eqnarray}
where $R_{2^n}(y) = {1\over 2} \; Q_{2^n}(y) + f_{2^n}(y) -\alpha \; h_{2^n}(y) $ and
\begin{eqnarray}
  Q_{2^{n}}(y) &=& b \left(f_{2^{n-1}}(b)  {\partial f_{2^{n-1}}(y-b)   \over \partial b} - f_{2^{n-1}}(y-b)  {\partial f_{2^{n-1}}(b)   \over \partial b}  \right) - f_{2^{n-1}}(b) \; f_{2^{n-1}}(y-b) \\
  f_{2^n}(y) &=& \int_0^{e_0} f_{2^{n-1}}(x)  \; Q_{2^{n-1}}(y-x) \; {\rm d}x \\
  h_{2^n}(y) &=& \int_0^{e_0}\; f_{2^{n-1}}(x) \; h_{2^{n-1}}(y-x) \; {\rm d}x
\label{qfh}
\end{eqnarray}
As in the previous cases,  the differential equation for the present case with $g=2^n+1$ can now be derived from eq.(\ref{dwg}) along with eq.(\ref{fg}). 

 The two independent solutions of the homogeneous part of eq.({\ref{dwg}) can be given as, with $\gamma=1+\alpha
 g$ and again using subscript $g$ instead of $2^n$,
\begin{eqnarray}
 f_{g,h}(y)=c_1 \;  \Gamma(1-\gamma) \; (g y)^{\gamma/2} \; I_{-\gamma}(2 \sqrt{gy}) +c_2 \; (-1)^{\gamma} \;  \Gamma(1+\gamma) \;(g y)^{\gamma/2} \; I_{\gamma}(2 \sqrt{gy})
\label{dwg1}
\end{eqnarray}
with $I_{\gamma}$ as the modified Bessel function and $c_1, c_2$ as the  constants of integration, to be determined from the boundary conditions. The above in turn gives the wronskin $w(x)$ of the two homogeneous solutions as $w(x)=(-g)^{(1+a g)} \;(1+a g) \; x^{a g}$. Following standard route,  the solution $f_{g,ih}$ of the inhomogeneous part of eq.(\ref{dwg}) can be written as
\begin{eqnarray}
 f_{g,ih}(y) &=& - \gamma_0 \; (g y)^{\gamma/2}  \left[A(g) \; I_{-\gamma}(2 \sqrt{gy})
 + B(g) \;  I_{\gamma}(2 \sqrt{gy}) \;  \right]
\label{dwg1}
\end{eqnarray}
with $\gamma_0=- (-1)^{\gamma} \; \Gamma(1+\gamma) \Gamma(1-\gamma) $ and
\begin{eqnarray}
A_g(y) &=&\int \; (g x)^{\gamma/2} \;  \; I_{\gamma}(2 \sqrt{gx}) \; R_g(x) \; (w(x))^{-1} \;{\rm d}x \\
B_g(y) &=& \int \; (g x)^{\gamma/2} \;  \; I_{-\gamma}(2 \sqrt{gx}) \; R_g(x) \; (w(x))^{-1} \; {\rm d}x.
\label{ab1} 
\end{eqnarray}

Using $f_g(y)=f_{g,h}(y)+f_{g,ih}(y)$, the general solution of eq.(\ref{dwg}) can now be written as 
\begin{eqnarray}
f_g(y) &=&  \left(c_1 \;  \Gamma(1-\gamma) - \gamma_0 \; A_g(y) \right)
 (g y)^{\gamma/2}  \; I_{-\gamma}(2 \sqrt{gy})  + \nonumber \\
 &&+ \left(c_2 \;  (-1)^{\gamma} \;  \Gamma(1+\gamma) - \gamma_0 \; B_g(y) \right) \; (g y)^{\gamma/2} \; I_{\gamma}(2 \sqrt{gy})
\label{fgw1}
\end{eqnarray}
 here again the constants $c_1, c_2$ can be detremined by imposing the boundary conditions at $y=0$ and $y=e_0$.  The substitution of $f_g(y)$ in eq.(\ref{rhogg}) then leads to {\it mdos} $\langle \rho_g \rangle$.

\subsection{Wishart Orthogonal ensemble (WOE)}

As in the previous case, the {\it sdos} $\langle \rho_{1}(e) \rangle$  in the edge region for  a WOE, can  still be described by eq.(\ref{rho1w}) with $f_1(x)$ given as \cite{fo1} 
\begin{eqnarray}
f_{1}(x)=f_{1,U}(x)  - { J_{a+1}(\sqrt{x}) \over 4 \sqrt{x}} \;\left( \int_0^{\sqrt{x}} \; J_{a+1}(v) \; {\rm d}v -1 \right) \label{wrho2}
\end{eqnarray}
where $f_{1,U}(x)$ is same as $f_1(x)$ of WUE case, given by eq.(\ref{fw})

In principle, proceeding as in previous cases, the exact differential equation for $f_g(y)$ for this case can again be derived. The derivation can however be simplified by noting that , near the lower edge $x \sim 0$, the $2nd$ term in eq.(\ref{wrho2}) is negligible with respect to first term and one can approximate
\begin{eqnarray}
f_1(x)  \approx f_{1,U}(x)  
\label{wrho3}
\end{eqnarray}
The above approximation in turn would again lead to eq.(\ref{dwg}) for {\it mdos} in WOE case. This indicates  the insensitiivty of the behaviour in the edge of the spectrum of a Wishart ensemble to exact symmetry conditions.

\section{Brownian ensembles}

A Brownian ensemble (BE) in general refers to an intermediate state of perturbation of a stationary random matrix ensemble by another one of a different universality class \cite{dy, fh, me, sp}.  The type of a BE, appearing during the cross-over, depends on the nature of the stationary ensembles and their different pairs may give rise to different BEs \cite{sp, psijmp}.  Here we consider the BEs appearing between stationary ensembles of Gaussian and Wishart type oly.

\subsection{Gaussian Brownian Ensembles}

A Brownian ensemble of Hermitian matrices $H$ can be described as $H=\sqrt{f} (H_0+t V)$  with $V(t)$ as a random perturbation of strength $t$, taken from a stationary Gaussian ensemble characterized by symmetry parameter $\beta$, and applied to an initial stationary state $H_0$ (see also \cite{psall}). Here $f=(1+\gamma \; t^2)^{-1}$ with $\gamma$ as an arbitrary positive constant.

The {\it sdos} for the above ensemble is described by a diffusion equation, referred as Dyson-Pastur equation, \cite{sp, psall}:
\begin{eqnarray}
{1\over 2\beta} \; {\partial \langle \rho_{1}\rangle \over\partial Y} &=&  {\partial \over \partial e} \left[ \gamma \; e  -  \alpha(e) \;  \right] \; \langle \rho_{1}\rangle 
\label{rr1}
\end{eqnarray}
with   $Y \propto {1\over 2 \gamma} \; \ln (1+ \gamma ; t^2)$ and
\begin{eqnarray}
\alpha(e) \equiv \alpha_e \equiv {\bf P}\int_{-\infty}^{\infty}   {\langle \rho_{1}(e')\rangle  \over e-e'}.
\label{alpe}
\end{eqnarray}
We note that here $e_0=- \sqrt{2N}$, the standard edge limit for Gaussian stationary ensembles (as eq.(\ref{rr1}) is derived for origin of the spectrum at $e=0$).

To derive $\langle \rho_{g}(e) \rangle$ in the edge regime for the present case, we again proceed  iteratively and start with $g=2$. Using the definition in eq.(\ref{mbd0}) for $\langle \rho_{2} \rangle$ and differentiating it with respect to $Y$, we have

\begin{eqnarray}
{\partial \langle \rho_{2} \rangle \over\partial Y} =  J_1 +J_2
\label{mbd5}
\end{eqnarray}
with 
\begin{eqnarray}
 J_1 &=& \int_{-\infty}^{e_0}  {\partial  \langle  \rho_{1}(x) \rangle  \over\partial Y} \; \langle \rho_{1}(e-x) \rangle ; {\rm d}x   \\
J_2 &=& \int_{-\infty}^{e_0} \langle  \rho_{1}(x) \rangle \; {\partial  \langle  \rho_{1}(e-x) \rangle  \over\partial Y} \; {\rm d}x .
\label{mbd6}
\end{eqnarray}
Substitution of eq.(\ref{rr1}) followed by partial integration reduces  $J_1$ as 
\begin{eqnarray}
 J_1 &=& S_2(e,e_0)   -\int_{-\infty}^{e_0}  \; (\gamma x - \alpha(x)) \langle  \rho_{1}(x) \rangle \; {\partial_{x}  \langle  \rho_{1}(e-x) \rangle} \; {\rm d}x\\
&=& S_2(e,e_0)   + \partial_{e} \int_{-\infty}^{e_0}  \; (\gamma x - \alpha(x)) \langle  \rho_{1}(x) \rangle \; \langle  \rho_{1}(e-x) \rangle \; {\rm d}x
\label{mbd6}
\end{eqnarray}
with $S_2(e,e_0) = (\gamma e_0 - \alpha(e_0)) \rho_1(e-e_0) \rho_1(e_0)$ and $\partial_{x} \equiv {\partial  \over\partial x}$.
Similarly,  following  substitution of eq.(\ref{rr1}) and 
writing ${\partial  \langle  \rho_{1}(e-x) \rangle  \over\partial (e-x)} = {\partial  \langle  \rho_{1}(e-x) \rangle  \over\partial e}$, $J_2$ can be rewritten as 
\begin{eqnarray}
 J_{2} &=&  \int_{-\infty}^{e_0}  \; \langle  \rho_{1}(x) \rangle \; {\partial_{e-x}}  (\gamma (e-x) - \alpha_{e-x}) \;  \langle  \rho_{1}(e-x) \rangle \; {\rm d}x  \\
&=& \partial_{e} \; \int  (\gamma (e-x) - \alpha_{e-x}) \; \langle  \rho_{1}(x) \rangle \; \langle  \rho_{1}(e-x) \rangle \; {\rm d}x  
\label{mbd6+}
\end{eqnarray}

Substitution of the above in eq.(\ref{mbd5}) gives 
\begin{eqnarray}
{1\over 2\beta} \;  {\partial \langle \rho_{2} \rangle \over\partial Y} &=&  S_2(e,e_0)  +  {\partial  \over\partial e}  \left(\gamma  e \langle  \rho_{2} \rangle -G_2(e) \right)   
\label{mbd7}
\end{eqnarray}
with $G_2(e)=\int_{-\infty}^{e_0} \left( \alpha_{x}+\alpha_{e-x} \right) \; \langle  \rho_{1}(x) \rangle \; \langle  \rho_{1}(e-x) \rangle \; {\rm d}x $. 
As the maximum contribution to $G_2$ comes from the neighborhood of $e_0$, it can be approximated as 
$G_2(e) \approx  \left( \alpha_{e_0} + \alpha_{e-e_0} \right) \int_{-\infty}^{e_0}  \; \langle  \rho_{1}(x) \rangle \; \langle  \rho_{1}(e-x) \rangle \; {\rm d}x =
\left( \alpha_{e_0} + \alpha_{e-e_0} \right) \langle  \rho_{2} \rangle$.
This reduces  eq.(\ref{mbd7}) as
\begin{eqnarray}
{1\over 2\beta} \;  {\partial \langle \rho_{2} \rangle \over\partial Y} 
 &=&  S_2(e,e_0) + {\partial\over \partial e} \left[\gamma e - \alpha_{2,e}\right]  \;   \langle \rho_{2}\rangle 
\label{mbd8}
\end{eqnarray} 
with $\alpha_{2,e} \equiv \alpha_2(e) =\alpha(e_0) + \alpha(e-e_0) $. Proceeding iteratively with eq.(\ref{mbd1}) again, one can similarly derive the diffusion equation for  $\langle \rho_{2^n} \rangle$:
\begin{eqnarray}
{1\over 2\beta} \;  {\partial \langle \rho_{2^n} \rangle \over\partial Y} 
 &=& \sum_{k=1}^n 2^{n-k} \; S_{2^{k-1}, 2^n-2^{k-1}} + {\partial\over \partial e} \left[\gamma e - \alpha_{2^n,e}\right]  \;   \langle \rho_{2^n}\rangle 
\label{mbd9}
\end{eqnarray}
with $S_{ab} = (\gamma e_0 - \alpha(e_0)) \rho_a(e_0) \rho_b(e-e_0)$ and 
$\alpha_{2^n,e} \equiv \alpha_{2^n}(e) =\alpha_{2^{n-1}}(e_0) + \alpha_{2^{n-1}}(e-e_0)$. 

We note that  the case $\alpha(e_0)=0$ implies $S_{ab}=0$ and eq.(\ref{mbd9}) for {\it mdos} reduces to the same form as that of {\it sdos} i.e eq.(\ref{rr1}).

\subsection{ Wishart Brownian ensembles}

 Consider an ensemble of $N_a \times N$  rectangular matrices $A(t)=\sqrt{f} (A_0+t V(t))$ with $f=(1+\gamma \; t^2)^{-1}$ \cite{sp, pslg} with $A_0$ as a fixed matrix and $\gamma$ as an arbitrary positive constant. The matrices   $L=A^{\dagger} A$ correspond to Wishart Brownian ensembles (WBE) if the matrices $V^{\dagger} V$ are taken from stationary Wishart ensembles e.g WOE or WUE. As clear, $A=A_0$ for $t\rightarrow 0$, $A \rightarrow V/ \sqrt{\gamma}$ for $t \rightarrow \infty$.

A variation of strength $t$ of the random perturbation $V$ leads to diffusion of the matrix elements $A_{kl}(t)=\sqrt{f} (A_{0;kl}+t V_{kl}(t))$. 
For $\rho_v(V) =\left(\frac{1}{2 \pi v^2}\right)^{\beta N_a N/2} {\rm e}^{-{1\over 2v^2} \; {\rm Tr} (V V^{\dagger})}$,  the diffusion equation for the {\it sdos} for the $L$-ensemble is described by a diffusion equation \cite{psall}:
\begin{eqnarray}
{1\over 2\beta} \; {\partial \langle \rho_{1} \rangle \over\partial Y} &=&  {\partial \over \partial e} \left( \gamma \; e  -   \alpha_w(e)  \right) \; \langle \rho_{1}\rangle 
\label{wr1}
\end{eqnarray}
with    $Y=-{1\over 2 \gamma} \; \ln f = {1\over 2 \gamma} \; \ln(1+ \gamma \; t^2)$ and $\alpha_w(e)= e \; \alpha(e)$ same as in eq.(\ref{rr1}).

We note that the form of the eq.(\ref{wr1}) is same as that of eq.(\ref{rr1}) except for $\alpha(e)$ in the latter replaced by $\alpha_w(e)$ in the former.
Thus proceeding again as in previous case,  
 the evolution of $\langle \rho_{2^n}\rangle$ can again be described by 
eq.(\ref{mbd9}) but with $\alpha(e) \to \alpha_w(e)$:
\begin{eqnarray}
{1\over 2\beta} \;  {\partial \langle \rho_{2^n} \rangle \over\partial Y} 
 &=& \sum_{k=1}^n 2^{n-k} \; S_{2^{k-1}, 2^n-2^{k-1}} + {\partial\over \partial e} \left[e (\gamma - \alpha_{2^n,e}) \right]  \;   \langle \rho_{2^n}\rangle 
\label{mbd10}
\end{eqnarray}
with $S_{ab} =e_0  (\gamma - \alpha(e_0)) \rho_a(e_0) \rho_{b}(e-e_0)$ and
$\alpha_{2^n,e} \equiv \alpha_{2^n}(e) =\alpha_{2^{n-1}}(e_0) + \alpha_{2^{n-1}}(e-e_0)$.

\subsection{Multiparametric Ensembles}

Consider an operator of a complex system  modeled by a multiparametric Gaussian ensemble of Hermitian matrices with probability  density

\begin{eqnarray}
\rho_{\mathcal H} ({\mathcal H},v,b)=C \; {\rm exp}[{-\sum_{k\le l} {1 \over 2 v_{kl}} ({\mathcal H}_{kl}-b_{kl})^2 }];
\label{rhog}
\end{eqnarray}
 here the variances $v_{kl}$ and mean values $b_{kl}$ can take arbitrary values (e.g. $v_{kl} \to 0$ for the non-random elements). As discussed in a series of studies \cite{psijmp, psand, psall}, the evolution of the {\it sdos} of the above ensemble is again described by eq.(\ref{rr1}) but now $Y$ is given by a combination of ensemble parameters, $Y= -{1\over  \gamma \; M}  \; \; {\rm ln}\left[ \prod_{k \le l} \; \prod_{q=1}^{\beta}|x_{kl}| \quad |b_{kl} + b_0|^2 \right] + constant$ and is referred as the ''complexity parameter''. Here $b_0 =1$ or $0$ if $b_{kl}=0$ or $\not=0$ respectively. 

The equivalence of the differential equation for {\it sdos} further implies that  {\it mdos} for a system, consiting of subunits statistically described by eq.(\ref{rhog}), is also given by eq.(\ref{mbd9}) however $Y$ is now the complexity parameter mentione above.

\section{Conclusion}

In the end, we summarize with our main results and open questions.

Based on the information about the ensemble averaged density of states of a single subunit of a many body system in its spectral edge, we derive the ensemble averaged many body density of states. Our analysis clearly indicates that the latter  changes with increasing number of subunits and differs from that of single body density of states even in non-interacting limit. 

While the analysis in the present study is confined to the edge of the spectrum, the derivation can directly be extended to any other energy range including bulk, within noninteracting approximation.  Although  the convolution route however is not applicable for the interacting many body systems, it would be interesting to compare some available results for the latter  with the results derived in the present study  and study the extent of their differences/ deviation and sensitivity to interaction parameters.

\acknowledgments
I am grateful to Professor Michael Berry for some technical help with {\it mathematica} code used for solving the differential equations. I also   thank {\it science and educational research borad (SERB), department of science (DST), India} for the financial support provided for the  research under MATRICES grant scheme.

\end{document}